\documentstyle[aps]{revtex}
\begin{document}
\begin{flushright}
SU-ITP-96-49\\
hep-th/ 9611094\\
November 12, 1996 \\
\end{flushright}
\vspace{1cm}
\begin{center}
\baselineskip=16pt

{\Large\bf  SUPERSYMMETRIC ROTATING BLACK HOLES AND ATTRACTORS }  \\

\vskip 2cm

{\bf Renata
Kallosh${}^{a}$, Arvind Rajaraman${}^{b}$  and
Wing Kai Wong${}^{a}$ \footnote{ E-mail:
kallosh@physics.stanford.edu,  arvindra@leland.stanford.edu,
wkwong@leland.stanford.edu}}
\\
 \vskip 0.8cm
${}^{a}$Physics Department, Stanford University, Stanford,   CA 94305-4060,
USA\\
\vskip .4cm
${}^{b}$Stanford Linear Accelerator Center,\\
    Stanford University, Stanford, California 94309, USA
\\

\vskip 1 cm

\end{center}
\vskip 1 cm
\centerline{\bf ABSTRACT}
\begin{quotation}

 Five-dimensional stringy rotating black holes are embedded into N=2
supergravity interacting with one vector multiplet.  The existence of an
unbroken
supersymmetry of the rotating solution is proved directly by solving the
Killing spinor equations.
The asymptotic enhancement of supersymmetry near the horizon in the
presence of
rotation is established via the calculation of the super-curvature.
The area of the horizon of the  rotating supersymmetric  black
holes is found to be $\sqrt {Z_{\rm fix}^{3 }- J^2}$, where $Z_{\rm fix}$ is
the
extremal value of the central charge in moduli space.

 \end{quotation}
\newpage

\section {Introduction}

It was shown by
 Tod \cite{Tod} that the Kerr-Newman metric in 4D
reaches the
supersymmetric limit when the mass is equal to the charge,  $m=|q|$,  for
arbitrary value of angular momentum $J$. However, the extreme limit
for this
solution is reached at $m^2-|q|^2=J^2$ and therefore the extremal
Kerr-Newman
solution does not have any unbroken supersymmetry. The
supersymmetric solution
with   $m=|q|$ for non-vanishing angular momentum $J$ turns out to
belong to the
Israel-Wilson-Perj\'{e}s family of metrics \cite{IWP} and is known
to have
naked singularities\cite{HaHa}. Other rotating black hole
solutions in 4D have been found with 1/2 and 1/4 of unbroken
supersymmetry of
N=4 theory but they always have naked
singularities (see \cite{BKO} for a review).

In contrast,  5D rotating extremal black hole
solutions have been constructed which have some unbroken
supersymmetry\cite{BMPV}. The
entropy of
these black holes as well as of the near extreme ones  was understood via the
counting of D-brane states in  \cite{BMPV} and in
\cite{BLMPSV}, respectively.

Extremal supersymmetric static black hole solutions behave as
attractors\cite{FKS};
moduli
take on fixed values at the horizon which depend only on the charges and not
on the values of the moduli at infinity. What is more, the area of these black
holes can
be found (without a knowledge of the metric) by extremizing the value of the
central
charge in moduli space \cite{FK1}. The extremal value $Z_{\rm fix}$ is then
related to the
area in
5D by
\begin{equation}
 A =   {\pi^2\over 3}Z ^{3/2}_{\rm fix}.
\label{area}\end{equation}
The precise relation is dimension-dependent; it was worked out for 4D
in\cite{FK1} and for
5D in \cite{CFGK}.

In this paper, we wish to study the analogues of these questions for
supersymmetric
rotating black holes. The above properties for static black holes are a
consequence of the fact that asymptotically close to the horizon, all the
supersymmetries are unbroken \cite{FK1}.
We will establish that the same phenomenon occurs for rotating black holes.
Using
this property, we can then show that

(a) there is still a fixed point for the moduli near the horizon, and

(b) the relation (1) is now modified to
\begin{equation}
 A =   {\pi^2\over 3}\sqrt{( Z_{\rm fix} ^{3}-J^2)}.
\label{newarea}\end{equation}

In section II, we will find the $N=2$ supergravity Lagrangian for
which the rotating black hole of \cite{BMPV} (henceforth referred to as the
BMPV black hole) is a solution. We will then
explicitly show that the supersymmetry transformation laws
are satisfied and that a Killing spinor exists. In the next section, we shall
examine the
integrability condition for the existence of Killing spinors. This condition is
trivial at the
horizon, indicating that it is satisfied for all spinors. This implies that we
have full
restoration of supersymmetry near the horizon. This will then imply the
properties (a) and (b)
above.

\section{Embedding of 5D rotating black holes into very special geometry}

The  general action for D=5, N=2 supergravity coupled to  N=2 vector multiplets
has been
constructed by  G\"unaydin,  Sierra  and  Townsend \cite{GST} ;  the
bosonic part
is referred to as very special geometry \cite{dWvP}.
The theory is completely defined by the prepotential
\begin{equation}
{\cal V} = {1\over 6} C_{IJK} X^I X^J X^K \ ,
\label{prep}\end{equation}
where  the  vectors
are labeled by $I=(0,i)$ where $i=1,\dots ,n$. This action corresponds to the
compactification of  11D supergravity down  to 5D
 on Calabi-Yau three-folds
\cite{Cadavid} with topological intersection form $C_{IJK}$.
The bosonic part of it is
\begin{eqnarray}
e^{-1} {\cal L} = -{1\over 2} R   - {1\over 4} G_{IJ}(\phi)  F_{\mu\nu} {}^I
F^{\mu\nu J}
 -{1\over 2} g_{ij}(\phi) \partial_{\mu} \phi^i \partial^\mu \phi^j
+{ e^{-1}  \over 48} \epsilon^{\mu\nu\rho\sigma\lambda} C_{IJK} F_{\mu\nu}^I
F_{\rho\sigma}^J A_\lambda^K  \ .
\label{veryspecial}\end{eqnarray}
The details and examples in the applications to 5D static black holes can
be found in \cite{CFGK}.
The simplest situation is  pure N=2 supergravity which has
 $n=0$ and only $C_{000}$  non-vanishing. This theory has no
scalars and no vectors besides the graviphoton.

The next simplest example of (\ref{veryspecial}) is
 the action of N=2 supergravity interacting with one vector multiplet in the
form presented in \cite{CFGK} \footnote{Here, as different from  \cite{CFGK},
we
will use the mostly minus metric and adapt our notation to this.}

\begin{equation}
e^{-1} {\cal L} = -{1\over 2} R   - {1\over 4} e^{{2\over 3}\phi} F_{\mu\nu} F
^{\mu\nu } - {1\over 4} e^{-{4\over 3}\phi} G_{\mu\nu} G ^{\mu\nu }
+{1\over 6} ( \partial_{\mu} \phi) ^2   -{ e^{-1}   \over 4
\sqrt2}\epsilon^{\mu\nu\rho\sigma\lambda} F_{\mu\nu}
F_{\rho\sigma  } B_\lambda \ .
\label{1v}\end{equation}
The supersymmetry  transformations of the fermionic fields with vanishing
fermions are
\begin{eqnarray}
\delta\psi_{\mu } &=& \nabla_\mu (\omega) \epsilon +{ 1\over 12 }
\Bigl(\Gamma_\mu{}^{\rho \sigma} - 4 \delta_\mu{}^ \rho  \Gamma^\sigma \Bigr)
(e^{{\phi \over
3}}
F_{\rho \sigma } - {1\over \sqrt 2} e^{-{2\over 3}\phi} G_{\rho \sigma})
\epsilon  \ ,\nonumber\\
\delta \chi&= & -{1\over 2\sqrt 3}\Gamma^\mu  \partial_\mu \phi \epsilon +
{1\over 4\sqrt
3} \Gamma ^{ \rho \sigma}\Bigl (e^{{\phi \over 3}} F_{\rho \sigma} +  \sqrt 2
e^{-{2 \phi
\over 3}} G_{\rho \sigma}
\Bigr)  \epsilon    \ .
\end{eqnarray}
The gravitational multiplet consists of the graviton $g_{\mu\nu}$ , gravitino
$\psi_{\mu }$ and the graviphoton field
$(e^{{\phi \over
3}}
F_{\rho \sigma } - {1\over \sqrt 2} e^{-{2\over 3}\phi} G_{\rho \sigma})
$. The vector multiplet includes  a scalar field $\phi$, a gaugino $ \chi$ and
the vector field of the vector multiplet $\Bigl (e^{{\phi \over 3}} F_{\rho
\sigma} +  \sqrt 2
e^{-{2 \phi
\over 3}} G_{\rho \sigma}
\Bigr) $.

Here we would like to find the rotating black hole with one half of unbroken
supersymmetry
in the theory of N=2 supergravity, interacting with one vector multiplet
. Up to a few rescalings  this is expected to be the rotating BMPV
black hole \cite{BMPV}. The ansatz for the metric is
\begin{eqnarray}
ds^{2}
= &&{\left(1 - { \mu \over{r^2}} \right)^2}
\left[dt - {{4J \sin^2\theta}\over{\pi (r^2-\mu)}} d\varphi
+ {{4 J \cos^2\theta} \over{\pi (r^2-\mu)}} d\psi \right]^2\nonumber \\
&&  - \left( 1 - {\mu\over{r^2}} \right)^{-2} dr^2
- r^2(d\theta^2 + \sin^2\theta d\varphi^2 + \cos^2\theta d\psi^2) \ .
\end{eqnarray}
The scalar field is a constant
\begin{equation}
e^{2 \phi}=  {8Q_F^2\over \pi^2 Q_H^2} = \lambda^6 \ .
\end{equation}
The gaugino equation for constant $\phi$ is then reduced to:
\begin{equation}
\delta \chi= {1\over 4\sqrt
3} \Gamma ^{ \rho \sigma}\Bigl (e^{{\phi \over 3}} F_{\rho \sigma} +  \sqrt 2
e^{-{2 \phi
\over 3}} G_{\rho \sigma}
\Bigr)  \epsilon =0 \ .
\end{equation}
It is satisfied by requiring that the vector field of the vector multiplet
vanishes, i.e.
\begin{equation}
G_{\rho \sigma}= - {1\over   \sqrt 2} e^{\phi} F_{\rho \sigma}  \  \
\Rightarrow  \qquad B =
- {{\lambda^3}\over{\sqrt{2}}} A \ .
\label{relation}\end{equation}
The vector field is
\begin{eqnarray}
A_t = {{\mu}\over{\lambda r^2}}\ , \qquad
A_\varphi=
                 {{4J \sin^2\theta}\over{\pi \lambda r^2}} \ , \qquad
A_\psi  = -
                 {{4J \cos^2\theta}\over{\pi \lambda r^2}}\ . \qquad
\label{vectansatz}\end{eqnarray}
For the mass and charges we have
\begin{equation}
M_{ADM} = {{3\pi\mu}\over{4}}
 \ , \qquad Q_H \equiv {{\sqrt 2 }\over{4\pi^2}} \int_{S^3}
{}^\star e^{-4\phi/3}G  = \mu/\lambda^2 \ , \qquad
Q_F \equiv {{\sqrt 2 }\over{16\pi}} \int_{S^3}
{}^\star e^{2\phi/3} F = -{{\pi}\over{2\sqrt{2}}}\mu\,\lambda \ .
\end{equation}

The relation between vector fields can be inserted into the gravitino
transformation:
\begin{equation}
\delta\psi_{\mu } = \nabla_\mu (\omega) \epsilon +{ 1\over 8 }
\Bigl(\Gamma_\mu{}^{\rho \sigma}  - 4 \delta_\mu{}^ \rho  \Gamma^\sigma \Bigr)
e^{{\phi \over
3}}
F_{\rho \sigma } = \nabla_\mu (\omega) \epsilon +{ \lambda \over 8 }
\Bigl(\Gamma^{\rho \sigma} \Gamma_\mu +2  \Gamma^\rho \delta_\mu{}^ \sigma
\Bigr)
F_{\rho \sigma }
\epsilon \ .
\end{equation}
It remains to show that the rotating black hole background admits 1/2
unbroken supersymmetry, i.e. that the gravitino equation has a zero mode. It is
useful at this stage to explain that our problem is actually reduced to the
problem of solving for a Killing spinor in the pure N=2 theory. Indeed, we may
use the relation between the vector fields in (\ref{relation})
as well as the fact that the scalar is a constant directly in the action
(\ref{1v}). We get
\begin{equation}
e^{-1} {\cal L} = -{1\over 2} R   - {3\over 8} \lambda^2  F_{\mu\nu} F
^{\mu\nu }   +{ e^{-1} \lambda^3  \over 8
}\epsilon^{\mu\nu\rho\sigma\lambda} F_{\mu\nu}
F_{\rho\sigma  } A_\lambda \ .
\label{v}\end{equation}
Upon rescaling the action and setting $\tilde F= \lambda {\sqrt 3\over 2} F$ we
can
rewrite the action as
\begin{eqnarray}
e^{-1} {\cal L}_5 = -{1\over 4} R(\omega)    - {1\over 4}  \tilde F_{\mu\nu}^2
+
{ e^{-1}  \over 6 \sqrt 3} \epsilon^{\mu\nu\rho\sigma\lambda}  \tilde
F_{\mu\nu}
\tilde F_{\rho\sigma}\tilde A_\lambda \ ,
\end{eqnarray}
and \begin{equation}
\delta\psi_{\mu } = \hat\nabla_\mu\epsilon= \nabla (\omega) \epsilon +{1 \over
4\sqrt 3 }
\Bigl(\Gamma^{\rho \sigma} \Gamma_\mu +2  \Gamma^\rho \delta_\mu{}^ \sigma
\Bigr)
\tilde F_{\rho \sigma }
\epsilon.
\label{kill}\end{equation}
This is the bosonic action of pure N=2 supergravity theory, presented in
\cite{Crem}.
The Killing equation (\ref{kill}) in the rotating black hole background has a
solution:
\begin{equation}
\epsilon = \sqrt{1-\frac{\mu}{r^2}}
\ e^{\frac{1}{2}\Gamma^4\Gamma^3\theta}
\ e^{\frac{1}{2}\Gamma^4\Gamma^1(\phi+\psi)}
\epsilon_0,
\end{equation}
where $\epsilon_0$ is a constant spinor satisfying
\begin{equation}
\label{constraint}
(1+\Gamma^0) \epsilon_0 = 0.
\label{constr}\end{equation}
The existence of this solution explicitly shows that we have
half of the supersymmetries unbroken in presence of rotation. Interestingly,
there is no trace of the rotation in the form of the Killing spinor. It has
been
observed before \cite{O1} that Killing spinors in spherical coordinates
may display a dependence on angles even if the geometry is spherically
symmetric.

We have also found that the Killing spinor in cartesian isotropic coordinates
is
simply
$$\epsilon = \sqrt{1-\frac{\mu}{r^2}}\epsilon_0$$
where the constant spinor  $\epsilon_0$ satisfies the constraint
(\ref{constr}).

\section{Enhancement of Unbroken Supersymmetry Near the Horizon}

For the study of the area-entropy formula in presence of rotation $J$ we would
like to consider the near horizon geometry. For all static supersymmetric
solutions near the horizon at $r\rightarrow r_0$  and one can exhibit the
$AdS_{2} \times
S^{3}$ geometry using  $\hat r = (r-r_0 )\rightarrow 0$ \cite{CFGK}
\begin{equation}
ds^2 = ({ 2\hat r \over r_0})^2 dt^2 - ({ 2\hat r \over r_0})^{-2}   d \hat
r^2 - r_0^2 d^2 \Omega_3  \ ,
\label{BR}\end{equation}
where the 3-sphere $S^{3}$ is defined  by
\begin{equation}
 r_0^2 d^2 \Omega_3 = r_0^2 (d\theta^2 + \sin^2\theta d\varphi^2 + \cos^2\theta
d\psi^2)  \ ,
\end{equation}
and the volume of this $S^{3}$,
\begin{equation}
A(J=0) = 2\pi^2 r_0^3,
\end{equation}
gives the area of the non-rotating black hole horizon in 5D.

For the rotating  solutions near the horizon  at $r^2 \rightarrow \mu \equiv
r_0^2 $ the metric
 does not split into the product space: there are non-diagonal components.
Apart from this, the $AdS_2$ part of the metric is the same but the 3-sphere is
distorted:
\begin{equation}
ds^2 =({ 2\hat r \over r_0})^2 dt^2 - ({ 2\hat r \over r_0})^{-2}   d \hat
r^2 -  {16 \hat r J \sin^2  \theta \over r_0^3 \pi } dt d\phi  +{16 \hat r J
\cos ^2  \theta \over r_0^3 \pi } dt d\psi  - r_0^2 d^2 \Omega_3 (J) \ ,
\label{rotBR}\end{equation}
and the metric for the distorted 3-sphere is
\begin{equation}
r_0^2 d^2 \Omega_3 (J) = r_0^2 \left (d^2 \Omega_3  - \left ({4J \over r_0^3
\pi }\right)^2 (\sin^2\theta d\varphi -\cos^2\theta d\psi)^2\right) \ .
\label{distort}\end{equation}
The volume of the distorted 3-sphere defines the area of the horizon of the
rotating black hole
\begin{equation}
A(J) = 2\pi^2 \sqrt {r_0^6 - J^2} \ .
\label{rotarea}\end{equation}
To clarify the relation of the radius of the distorted 3-sphere $r_0$ to the
minimum of the central charge, we need to study the supersymmetry near the
horizon and find out whether
it is enhanced as in the non-rotating case.

The integrability condition for the existence of the Killing spinor  defines
the super-curvature as

\begin{equation}
[\hat\nabla_a,\hat\nabla_b] \epsilon =  \hat R_{ab}\ \epsilon = 0.
\label{supercurv}\end{equation}

If this is satisfied for arbitrary $\epsilon$, we have fully
restored supersymmetry.  If a constraint on $\epsilon$ is needed,
supersymmetry is partially broken.

We will find that supersymmetry is asymptotically restored near the
horizon and half broken away from the horizon.

In fact, the integrability condition takes the form

\begin{equation}
\label{integ}
\hat R_{ab} = \frac{(r^2- r_0^2 )}{r^6} X_{ab} (1+\Gamma^0),
\end{equation}
where $X_{ab}$ will be given  below.

{}From this equation we can see that $\hat R_{ab}$ approaches
zero near the horizon and so the integrability condition is
satisfied for arbitrary $\epsilon$, which implies supersymmetry
is asymptotically fully restored.

Away from the horizon, when the constraint (\ref{constraint}) is imposed,
the integrability condition is also satisfied, signifying half
unbroken supersymmetry, consistent with the results of the previous section.

The explicit form of $X_{ab}$ in  (\ref{integ}) is
\begin{eqnarray}
X_{ab} &=&  X^0_{ab} + \frac{4J }{\pi r} X^1_{ab} \nonumber \\
X^0_{ab} &=& r_0^2 \eta \Gamma_{ab},
\end{eqnarray}

where
\begin{equation}
\eta =
\begin{array}{cl}
1 &(a,b) = (1,2), (1,3) \;  \text{or}\ (1,4) \\
3 &(a,b) = (0,1)\\
-1 &\text{otherwise}.
\end{array}
\end{equation}

The nonzero terms in $X^1_{ab}$ are
\begin{eqnarray}
X^1_{13}&=&-X^1_{24}=+4\sin\theta\Gamma^1-2\cos\theta\Gamma^2\\
X^1_{14}&=&+X^1_{23}=-4\cos\theta\Gamma^1-2\sin\theta\Gamma^2\\
X^1_{12}&=&+X^1_{34}=+2\cos\theta\Gamma^3+2\sin\theta\Gamma^4
\end{eqnarray}
with the corresponding antisymmetric parts.

The fact that the super-curvature
near the horizon vanishes even in the presence of rotation allows us to extend
the arguments about the universality of the area-entropy formula of static
black
holes to the rotating ones.

For the most general 5D rotating supersymmetric black hole we have to study
both the supercurvature and the gaugino supersymmetry transformations. Using
the very
special geometry language we have for the theory of N=2 supergravity,
interacting with $n$ vector multiplets
\begin{equation}
\delta \lambda _i = - {i\over 2} g_{ij}(\phi) \Gamma^\mu  \partial_\mu  \phi^j
\epsilon  +  {1\over 4}
\left({3\over 4}\right)^{2/3} t_{I,i}   \Gamma^{\mu\nu}  F_{\mu\nu}^I
\epsilon.
\end{equation}
The doubling of unbroken supersymmetry near the horizon is possible only under
the
condition that there is a fixed point where all $n$ the scalars have vanishing
derivatives
\begin{equation}
e_a{}^\mu  \partial_\mu  \phi^j=0
\end{equation}
and simultaneously the vector fields of the vector multiplets vanish
\begin{equation}
t_{I,i}   \Gamma^{\mu\nu}  F_{\mu\nu}^I=0 \qquad \Longrightarrow  \qquad
\partial_i Z =0.
\end{equation}
Hence the moduli take on fixed values at the horizon, and the central charge is
extremized.
A particular example of this is eq. (\ref{relation}) which defines the fixed
scalar in terms of the charges and minimizes the central charge.
For a static black hole, we could then, following \cite{FK1}, deduce the
area of the black hole to be $A= {\pi^2\over 3}\sqrt {Z_{\rm fix}^{3 }}$.
However,
as seen in equation (\ref{rotarea}), the rotating black hole has a modified
area.
 This modifies the relation to
\begin{equation}
A= {\pi^2\over 3}\sqrt {Z_{\rm fix}^{3 }- J^2}.
\end{equation}
We have found that it is  correct for all known rotating solutions, including
those with scalars changing between infinity and the horizon \cite{Tsey,CY} and
may be considered as a prediction for rotating solutions which may be found
later.

\section {Acknowledgements}
We are grateful to R.Myers, T.~Ort\'{\i}n and  A. Peet, for discussions.
The work of R.K. and W.K.W. is supported by the  NSF grant
PHY-9219345. The work of A.R. is supported in part by the Department of Energy
under contract no. DE-AC03-76SF00515.

%\appendix
\section*{Appendix}
Here are some conventions, definitions and representations we used.

Greek letters denote curved space indices, roman letters denote
inertial frame indices.

\begin{equation}
\eta_{ab} = \text{diag}(+1,-1,-1,-1,-1) \\
\end{equation}
\begin{equation}
\Gamma^0 = \left[
\begin{array}{cccc}
 0 &  0 &  0 & -i \\
 0 &  0 &  i &  0 \\
 0 & -i &  0 &  0 \\
 i &  0 &  0 &  0
\end{array} \right]
\qquad \Gamma^1 = \left[
\begin{array}{cccc}
 i &  0 &  0 &  0 \\
 0 & -i &  0 &  0 \\
 0 &  0 &  i &  0 \\
 0 &  0 &  0 & -i
\end{array} \right]
\qquad \Gamma^2 = \left[
\begin{array}{cccc}
 0 &  0 &  0 &  i \\
 0 &  0 & -i &  0 \\
 0 & -i &  0 &  0 \\
 i &  0 &  0 &  0
\end{array} \right]
\end{equation}
\begin{equation}
\ \Gamma^3 = \left[
\begin{array}{cccc}
 0 & -i &  0 &  0 \\
-i &  0 &  0 &  0 \\
 0 &  0 &  0 & -i \\
 0 &  0 & -i &  0
\end{array} \right]
\qquad \Gamma^4 = \left[
\begin{array}{cccc}
 0 &  1 &  0 &  0 \\
-1 &  0 &  0 &  0 \\
 0 &  0 &  0 & -1 \\
 0 &  0 &  1 &  0
\end{array} \right]
\end{equation}
\begin{equation}
\nabla_\mu \epsilon =
\partial_\mu \epsilon +
\frac{1}{4} \omega_{\mu ab}
\Gamma^{ab}\epsilon
\end{equation}
\begin{equation}
[\hat\nabla_a,\hat\nabla_b] =
[\nabla_a, \nabla_b] +
\nabla_{[a}G_{b]}+
[G_a, G_b],
\end{equation}
where
\begin{equation}
G_a ={1 \over
4\sqrt 3 } (\Gamma^{cd}\Gamma_a+2\Gamma^c \delta^{d}_a ) \ F_{cd}.
\end{equation}

\end{document}